\documentclass[twocolumn,english,conference]{IEEEtran}
\usepackage[T1]{fontenc}
\usepackage{xcolor}
\usepackage{babel}
\usepackage{amsthm}
\usepackage{amsmath}
\usepackage{graphicx}

\makeatletter

\providecommand{\tabularnewline}{\\}

\theoremstyle{plain}
\newtheorem{thm}{\protect\theoremname}
\theoremstyle{definition}
\newtheorem{defn}[thm]{\protect\definitionname}

\usepackage[caption=false,font=footnotesize]{subfig}

\makeatother

\providecommand{\definitionname}{Definition}
\providecommand{\theoremname}{Theorem}

\begin{document}
\global\long\def\var{\mathrm{var}}

\newcommand{\ER}{Erd\"{o}s-R\'{e}nyi }

\newcommand{\BA}{Barab\'{a}si-Albert}

\title{Coding of Graphs with Application to Graph Anomaly Detection}

\author{\IEEEauthorblockN{Anders H{\o}st-Madsen and June Zhang}\IEEEauthorblockA{Department of Electrical Engineering, University of Hawaii, Manoa\\
Honolulu, HI, 96822, Email: \{ahm,zjz\}@hawaii.edu}}
\maketitle
\begin{abstract}
This paper has dual aims. First is to develop practical universal
coding methods for unlabeled graphs. Second is to use these for graph
anomaly detection. The paper develops two coding methods for
unlabeled graphs: one based on the degree distribution, the second
based on the triangle distribution. It is shown that these are efficient
for different types of random graphs, and on real-world graphs. These
coding methods is then used for detecting anomalous graphs, based
on structure alone. It is shown that anomalous graphs can be detected
with high probability.
\end{abstract}

\section{Introduction}

A popular research problem in data mining is graph anomaly detection, which has applications in areas ranging from finance to power grid operation to detecting social trends \cite{noble2003graph,eberle2007anomaly,akoglu2015graph}.  In this paper we explore using description
length for graph anomaly detection; that is, we encode the graph using
a lossless source coder, and use the resulting codelength as the decision
criteria. While minimum description length (MDL) has been used in
the connection with graph anomaly detection, the application has only
been for model selection in time-series analysis. As far as we know, this paper
is the first to consider using description length directly for anomaly
detection.

Reference \cite{ChoiSzpankowski12} was the first paper to develop practical source coding
algorithms for graphs.  To use source coding for description length analysis, the codelength has
to reflect the information in the graph, and the only information \cite{ChoiSzpankowski12} reflects really is the edge probability $p$ (see discussion later). This paper therefore develops new practical
(universal) source coding algorithms based on more informative statistics. This focus is different than other recent papers
in graph coding \cite{DelgoshaAnantharam17,LuczakSzpankowski17, LuczakSzpankowski17b, AsadiAbbeVerdu17} that are aimed more at entropy analysis.

\subsection{Graphs}

The structure of a graph is defined by the set of \textbf{vertices}
(also called nodes) $\mathcal{V}$, and the set of \textbf{edges},
$\mathcal{E}$. Usually, the ordering of the vertices are irrelevant, and in that
case we call the graph \textbf{unlabeled}; we will only consider unweighted, unlabeled, undirected graphs in this paper.
A graph, $G(\mathcal{V},\mathcal{E})$, is often represented by the
\textbf{adjacency matrix}, $\mathbf{A}=[A_{ij}]$, a $|\mathcal{V}|\times|\mathcal{V}|$
matrix where $A_{ij}=1$ if $(i,j)\in\mathcal{E}$. The degree of a vertex is the number of edges emanating from the vertex.
The \textbf{degree distribution} is the collection of the degrees
of all the nodes in the graph and is an often used statistics to differentiate between different classes of random graphs such as \ER\, \BA\, or Watts-Strogatz graphs \cite{barabasi2016network}. There is a one-to-one correspondence between binary, symmetric matrices
and unweighted, undirected graphs, and coding of graphs is therefore equivalent to coding
binary, symmetric matrices.

%


\subsection{Description Length}

The description length of the data is the number of bits required
to describe the data exactly: the data is turned into a stream of
bits, and from this the data should be able to be recovered exactly
by a decoder. We are only concerned with the length of the encoding, i.e.,
the number of bits output be encoder.

The central idea here is that the description length has some relationship
with the \textquotedbl{}meaning\textquotedbl{} of data. For example,
Rissanen considered \textquotedbl{}useful information\textquotedbl{}
in \cite{Rissanen86b}. More concretely, description length can be
used for data analysis. A traditional application, in particular in
terms of minimum description length (MDL) \cite{Rissanen83},
has been for model selection in data analysis.
The methodology
we will develop for graph coding can also be used for model selection
for more general data sets. However, we are more interested in description
length as a general data processing tool beyond simple model selection.
One example is atypicality which is described in Section \ref{Anomaly.sec}.

A central principle of description length is the constraint that a
decoder should be able to reconstruct the original data from an (infinite)
stream of bits. One manifestation is of course the Kraft inequality
\cite{CoverBook}, but the principle is more general. Since most source
coding algorithms are sequential, decodability then means that the
decoder can only use past decoded information to decode future data.
For graphs, this is much more complicated to satisfy than for sequences. Decodability
now becomes an algorithmic constraint rather than a probabilistic
one, moving description length theory closer to Kolmogorov complexity
\cite{LiVitanyi,CoverBook}. 


\section{\label{Coding.sec}Coding}

We will base graph coding on the adjacency matrix \textendash{} due
to symmetry, only the lower triangular part has to be coded. However,
usually the numbering of nodes is irrelevant. The resulting graph
modulo automorphisms is called the structure \cite{ChoiSzpankowski12}.
Using this in encoding can lead to smaller codelength. Importantly,
for data analysis, clearly the structure is more relevant, and description
length therefore should be based on the structure.

\begin{table}[tbh]
\begin{centering}
\begin{tabular}{cccc}
1 & \textcolor{blue}{1} & \textcolor{red}{1} & $\cdots$\tabularnewline
1 & \textcolor{blue}{1} & \textcolor{red}{1} & $\cdots$\tabularnewline
1 & \textcolor{blue}{1} & \textcolor{red}{0} & $\cdots$\tabularnewline
1 & \textcolor{blue}{0} & \textcolor{cyan}{1} & $\cdots$\tabularnewline
1 & \textcolor{blue}{0} & \textcolor{cyan}{0} & $\cdots$\tabularnewline
0 & \textcolor{green}{1} & \textcolor{purple}{1} & $\cdots$\tabularnewline
0 & \textcolor{green}{1} & \textcolor{purple}{0} & $\cdots$\tabularnewline
0 & \textcolor{green}{0} & \textcolor{pink}{1} & $\cdots$\tabularnewline
0 & \textcolor{green}{0} & \textcolor{pink}{0} & $\cdots$\tabularnewline
\end{tabular}
\par\end{centering}
\caption{\label{Stein.tab}The first column has one group, the second two (blue/green),
the third four (red/cyan/purple/pink).}
\vspace{-0.3in}
\end{table}
The adjacency matrix is a binary matrix, and coding this is therefore
similar to the problem considered by Steinruecken in \cite{Steinruecken15},
on which we will base our coding. Steinruecken considered coding of
unordered iid sequences, which we will think of as a matrix. We can
state the approach more abstractly as follows: we first sort the rows
according to some criterion (e.g., lexicographically). The coding
is done on the sorted matrix, and only the sorted matrix is reproduced
(exactly) at the receiver. The trick is to sort in such a way that
coding of the sorted matrix is more efficient than coding the original
matrix. The procedure in \cite{Steinruecken15} is to first sort the
sequences lexicographically (with 1 coming before 0). We say that
the sequences are grouped: the first group is all sequences, the next
two groups are sequences that start with 1/0, which is then subgrouped
into sequences starting with 11/10/01/00, see Table \ref{Stein.tab}.
An efficient code is as follows: we first transmit the number of ones
in the first column (the first group). The next column is divided
into two groups: those rows that has 1 in the first column, and those
that have 0. We transmit the number of ones in each group. When the
sequences are iid, the number of ones is binomially distributed, which
can be used for encoding. We continue this way (with empty groups
not encoded). 

This approach can also be applied to adjacency matrices, with the
modification that when we permute the rows during sorting, we have
to do the same permutation of columns to preserve symmetry. This turns
out to be equivalent to the algorithm in \cite{ChoiSzpankowski12},
but describing it this way reveals that the approach in \cite{ChoiSzpankowski12}
is strongly aimed at \ER graphs. From a data analysis point of view
this is problematic. The only parameter the algorithm in \cite{ChoiSzpankowski12}
is sensitive to is the average node degree $\bar{k}$ (equivalently
$p$). Consider anomalous graph detection in terms of atypicality
(this is described in more detail in Section \ref{AnomolousGraph.sec}):
We compare the codelength of encoding the graph with a given learned
coder and a universal coder. Since the only parameter \cite{ChoiSzpankowski12}
is sensitive to is $p$, this corresponds to a hypothesis test of
$p=p_{0}$ versus $p\neq p_{0}$. This is not irrelevant, but it is
far from what we do with sequences, where we can test a given FSM
against the whole class of alternative FSM. Thus, to be effective
for data analysis, we need much more effective coders. In the following
we will describe two such coders.

\subsection{\label{Degree.sec}Coding Using Degree Distribution}

Assume we know the degree distribution $P(k)$, either from a model,
from learning, or from the given graph. How can we take this into
account in coding? Consider coding of a given column of the sorted
adjacency matrix, as outlined above. Important here is what the decoder
already knows, from previous columns: it knows the number of ones
above the diagonal, it knows the number of groups $g$, and it knows
the size $s_{i}$ of each group; let $s=\sum_{i=1}^{g}s_{i}$. We
first encode the (total) degree of the node. Call the number of ones
above the diagonal $\bar{k}$. We can use the coding distribution
\begin{equation}
P(k|k\geq\bar{k})=\frac{P(k)}{\sum_{j=\bar{k}}^{\infty}P(j)}\label{Pk.eq}
\end{equation}
The decoder now has encoded the number of new ones (or edges) to encode.
The encoder needs to encode which configuration of the $k-\bar{k}$
is seen; that is, how many ones $k_{i}$ are in each group, subject
to the total count being $k-\bar{k}$. We assume that every sequence
with $k-\bar{k}$ ones is equally likely, so calculating the probability
of seeing a specific configuration is just a counting problem. In
total there are $\left(\begin{array}{c}
s\\
k-\bar{k}
\end{array}\right)$ sequences with $k-\bar{k}$, and there are $\left(\begin{array}{c}
s_{i}\\
k_{i}
\end{array}\right)$ ways to arrange the $k_{i}$ ones in each group. The coding probability
of a specific configuration therefore is
\[
\log P=\sum_{i-1}^{g}-\log\left(\begin{array}{c}
s_{i}\\
k_{i}
\end{array}\right)+\log\left(\begin{array}{c}
s\\
k-\bar{k}
\end{array}\right)
\]

A central assumption here is that at time of decoding a given column,
the decoder knows the number of ones $\bar{k}$ above the diagonal
so that it can calculate (\ref{Pk.eq}). This is satisfied if the
rows and columns are first sorted lexicographically, which can be
seen as follows. Suppose $i$ columns have been coded/decoded. The
decoder knows the first $i$ columns and rows in the (sorted) adjacency
matrix: this is clearly possible to reconstruct from the number of
ones in each group until column $i$ and the fact of the sorting.
The next row is chosen by the encoder among those among the remaining
$n-i$ columns that has highest sort order based on the first $i$
columns. No matter which column is chosen, the decoder knows the first
$i$ bits, and therefore the number of ones above the diagonal. 

It is not necessary to explicitly sort the adjacency matrix. Instead
one can use the same partitioning algorithm from \cite{ChoiSzpankowski12}.
While not very explicit in the paper, they actually sort the adjacency
matrix in the way they choose the next node to encode. It is seen most
clearly from \cite[Fig. 3]{ChoiSzpankowski12}.

\subsection{Coding of Triangles}

Edges are the most fundamental building block of graphs. A more complex
building block is triangles, i.e., a cycle graph with three nodes,
which is also a 3-clique. Statistics about triangles are often used
to characterize graphs \cite{barabasi2016network}. One statistic is the following. Consider
three connected nodes $i\leftrightarrow j\leftrightarrow k$; we let
$p_{\triangle}$ be the probability that there is also an edge $i\leftrightarrow k$.
We can use this for coding as follows. Let the current node to be
coded be node $i$, and suppose we want to code whether or not there
is an edge to node $k$. We now look for a common neighbor $j$ of
nodes $(i,k)$ \emph{among nodes already coded}. If such a node exists,
we encode the edge $i\leftrightarrow k$ using $p_{\triangle}$; otherwise,
we use $p$. This can be used together with the structure encoding
of Table \ref{Stein.tab}: Notice that all groups have exactly the
same connections to prior encoded nodes. Thus all the nodes $k\in G$
in a group either has a common previously encoded neighbor with node
$i$, or none have. Therefore, they can all be encoded with either
$p_{\triangle}$ or $p$. That is, the number of ones in the group
can be encoded with a binomial distribution with probability either
$p_{\triangle}$ or $p$.

\subsection{\label{Calc.sec}Calculation and Encoding of Statistics}

We consider encoding in two scenarios: learned coding, where we are
given a set of training graphs and have to learn the statistics; this
statistics is known both by encoder and decoder. Second, universal
coding, where the encoder encodes a single graph and also has to communicate
to the decoder what is the statistic.

For learned coding, the edge probability $p$ can be estimated straightforwardly
as an average. The degree distribution is estimated through a histogram.
To estimate $p_{\triangle}$ is more tricky. We select randomly three
connected nodes $i\leftrightarrow j\leftrightarrow k$ and calculate
$p_{\triangle}$ as a an average. However, the value of $p_{\triangle}$
depends on how the nodes are selected. When $p_{\triangle}$ is used
for coding, the triple of nodes is chosen in a specific way. The best
estimate is therefore found by performing the coding on the training
graphs. Notice that in that case the edges are divided into those
coded with the triangle probability $p_{\triangle}$ and those coded
with $p$. However, those edges not (coded) in a triangle could be
special. Instead of using the general $p$, we could estimate that
$p$ directly; we call this $\check{p}_{\triangle}$. In general $p\neq\check{p}_{\triangle}$,
but in many cases they are very close.

For universal coding, there are two possible approaches, best outlined
in \cite[Section 13.2]{CoverBook}: the encoder can estimate the parameters
of the coding distribution and inform the decoder of the estimate.
Or, the coding distribution can be sequentially calculated. For encoding $p$ for iid coding
the two approaches are essentially equivalent. The number of bits
required to encode the number of ones is about $\log\frac{n(n-1)}{2}\approx2\log n$
bits. For the degree distribution, we calculate the degree histogram
for the whole graph, and use this for coding. The degree of a node
is between 0 and $n-1$. We can therefore think of the degree histogram
as putting each of the $n$ (unlabeled) nodes into one of $n$ buckets,
and encoding this can be done by encoding the counts in the buckets.
The number of possible configurations is a standard problem in combinatorics:
$\left(\begin{array}{c}
2n-1\\
n
\end{array}\right)$, which can be transmitted with $\log\left(\begin{array}{c}
2n-1\\
n
\end{array}\right)=nH\left(\frac{n}{2n-1}\right)+\frac{1}{2}\log\frac{2n-1}{n^{2}}+c\approx n-\frac{1}{2}\log n$ bits ($|c|\leq2)$ . Of course, there is a relationship between the
degrees of nodes in the graph, and if we took this into consideration,
it might be possible to encode the degree histogram slightly more
efficient.

For triangle coding, we use sequential estimation of $p_{\triangle}$
and $\check{p}_{\triangle}$, specifically the KT estimator \cite{KrichevskyTrofimov81,WillemsAl95},
which is
$
\hat{p}=\frac{n_{1}+\frac{1}{2}}{n_{1}+n_{0}+1}
$,
where $n_{1},n_{0}$ is the number of ones and zeros seen previously.
The probabilities $p_{\triangle}$ and $\check{p}_{\triangle}$ are
not updated after each bit, but rather after each group is encoded.

\subsection{Numerical Results}

\begin{figure}[tbh]
\begin{centering}
\includegraphics[width=3.5in]{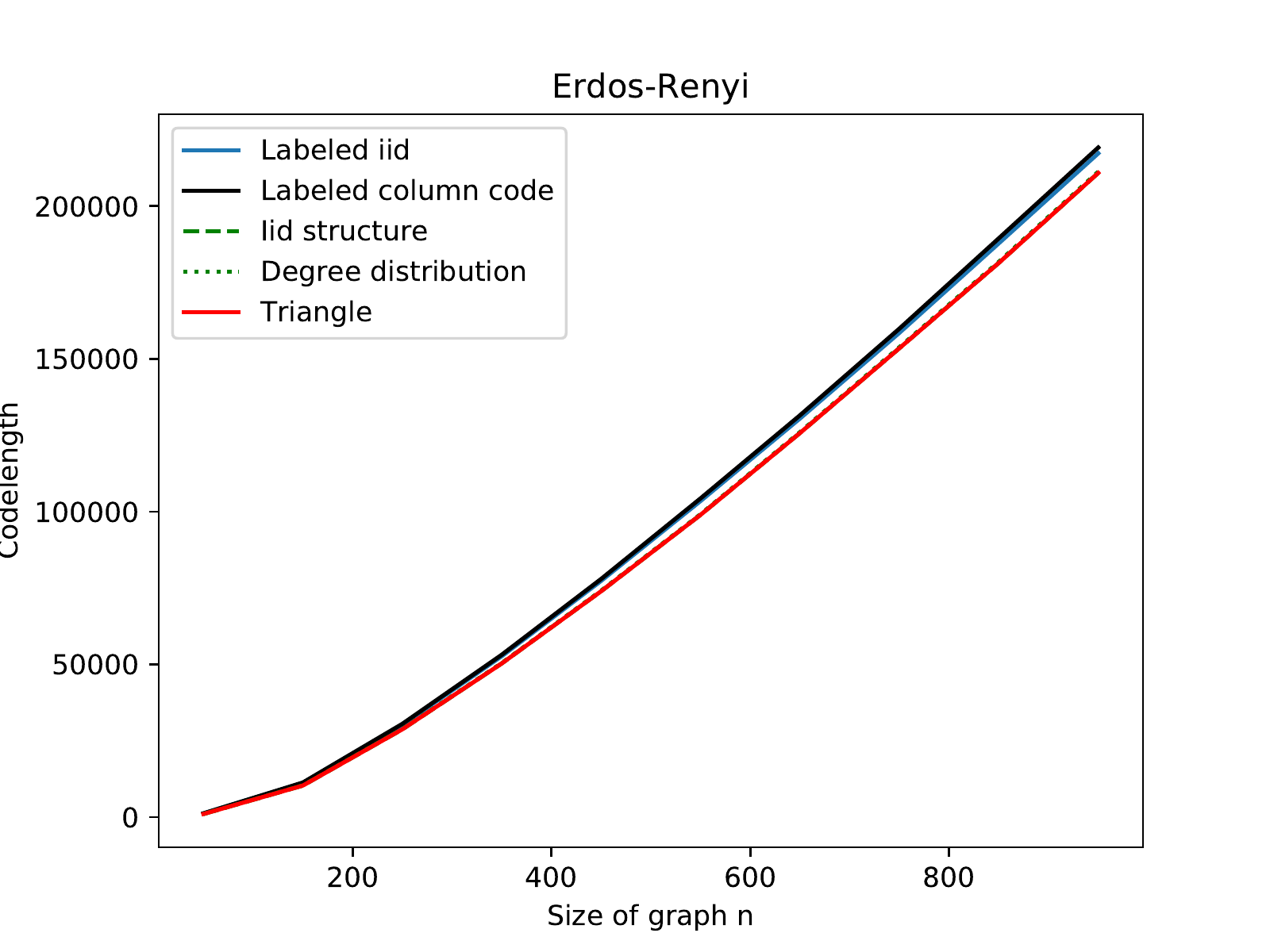} 
\par\end{centering}
\caption{\label{ERcomp.fig} Comparison of different codelengths for a ER graph
with $p=\min\{\frac{1}{2},\frac{100}{n}\}$}
\vspace{-0.2in}
\end{figure}

\begin{figure}[tbh]
\begin{centering}
\includegraphics[width=3.5in]{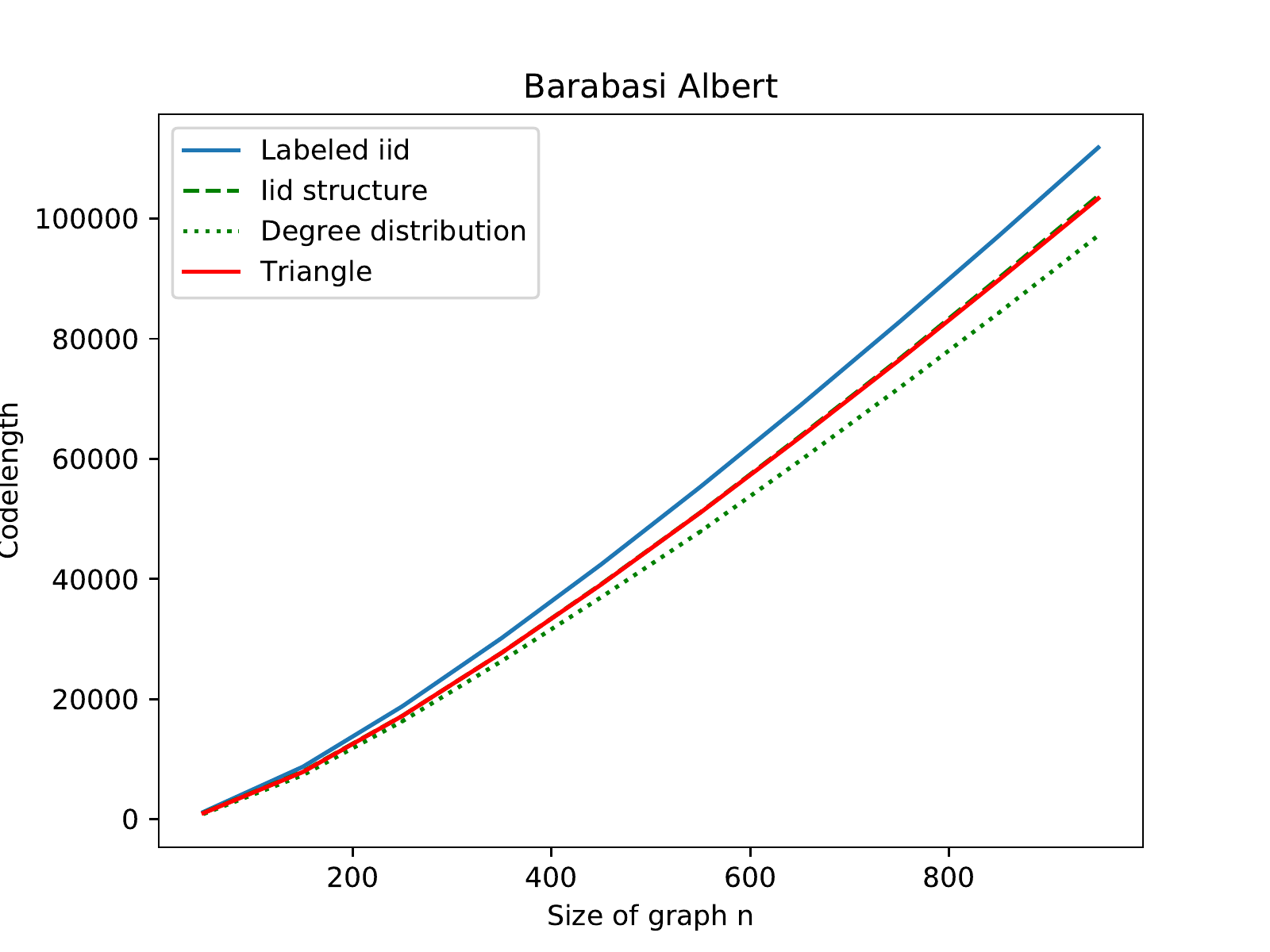} 
\par\end{centering}
\caption{\label{GraphCodeComp.fig} Comparison of different codelengths for
a BA graph with $m=20$.}
\end{figure}

\begin{figure}[tbh]
\includegraphics[width=3.5in]{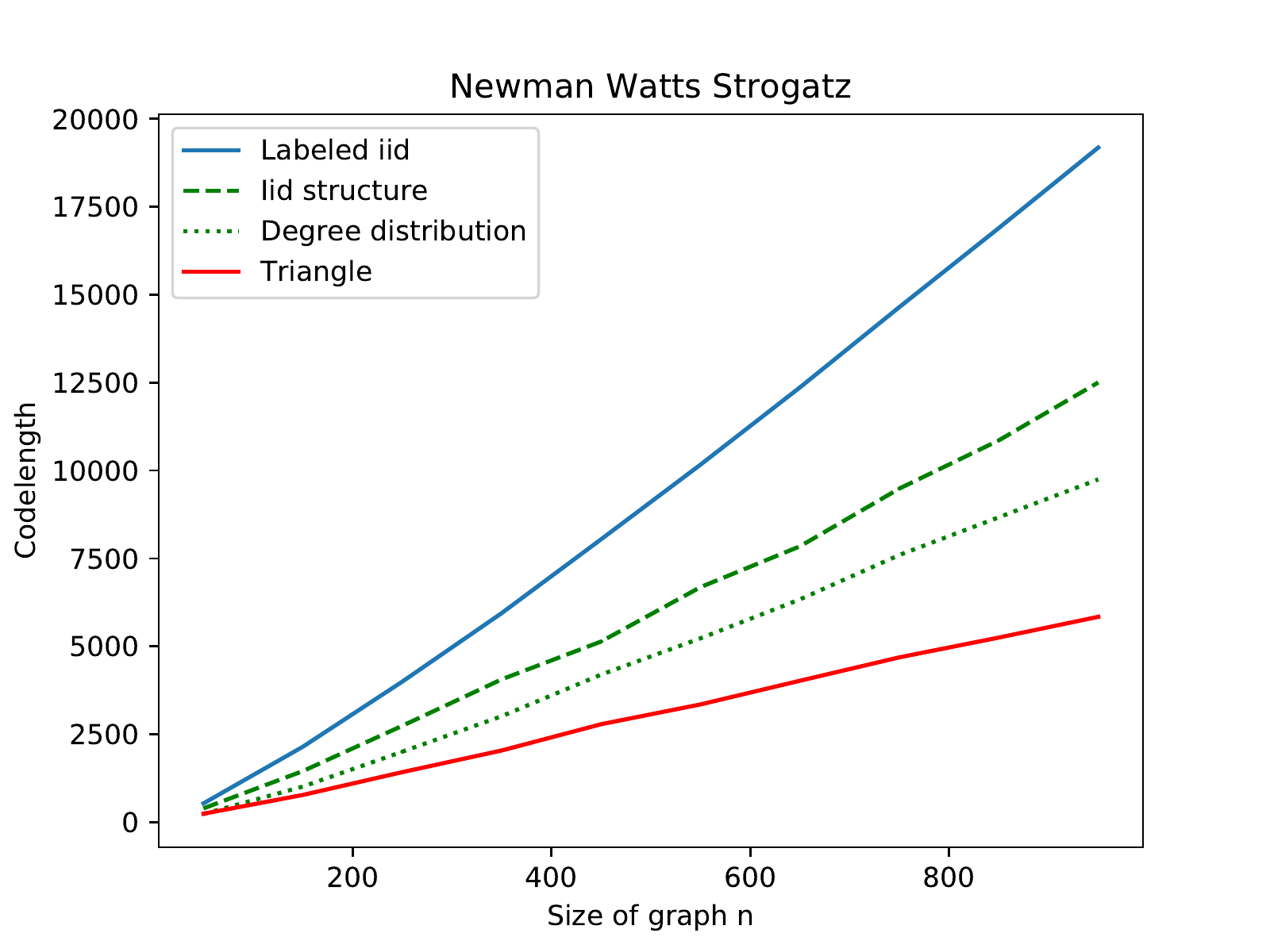}

\caption{\label{nws.fig}Comparison of different codelengths for a  Watts Strogatz graph \cite{watts1998collective}
with $k=5$ and $p=0.1$.}
\vspace{-0.2in}
\end{figure}

Some results can be seen in Fig. \ref{ERcomp.fig}-\ref{nws.fig}.
In all cases, learning was done on 50 graphs prior to coding. For
\ER  graphs, the iid structure code is most efficient, but all structure
codes give about the same codelength. For \BA\  graphs, coding using
the degree distribution is most efficient, and for Newman Watts Strogatz 
graphs \cite{watts1998collective}, using the triangle probability is most efficient. This shows
that there is no single efficient code for all graph structures.

We also did some experiments on real-world graphs, both obtained from \cite{Davis11}. For those graphs
there is no training, so the universal coding is needed. For both
graphs, using degree distribution is most efficient. However, transmitting
the degree histogram is expensive, and considering that, the triangle
coding is most efficient. In light of this one could consider better
ways to represent the degree distribution (e.g., a parametric representation),
but we have not explored that. 

\begin{table}[tbh]
\begin{centering}
\begin{tabular}{|c||c|c|}
\hline 
 Codelength $\searrow$  & Protein graph & Power graph\tabularnewline
\hline 
\hline 
Labeled iid & 20513 & 81077\tabularnewline
\hline 
Structure iid & 8796 & 32013\tabularnewline
\hline 
Degree distribution & 7290 & 27651\tabularnewline
\hline 
Degree distribution with overhead & 8743 & 32586\tabularnewline
\hline 
Triangle & 8369 & 26507\tabularnewline
\hline 
\end{tabular}
\par\end{centering}
\caption{\label{RealGraph.tab}Real-world graphs. The protein graph is the largest connected component of a network of protein interactions in the yeast Saccharomyces cerevisiae. The power graph represents the US Western power grid.}
\vspace{-0.3in}

\end{table}

\section{\label{Anomaly.sec}Anomaly Detection}

For detecting anomalous graphs, we will use atypicality developed
in \cite{HostSabetiWalton15}, which is described by

\begin{defn}
\label{atypdef.thm}A sequence is atypical if it can be described
(coded) with fewer bits in itself rather than using the (optimum)
code for typical sequences. 
\end{defn}
The papers \cite{HostSabetiWalton15,Host16BD} show that atypicality
has many desirable theoretical properties and that it works experimentally for
sequences. Specifically for anomaly detection, the paper \cite{Host16BD}
shows that atypicality is (asymptotically) optimum for finite state
machine (FSM) We will say that two FSM are \emph{distinct} if they
have no identical classes.  Then
\begin{thm}[\cite{Host16BD}]
\label{FSMoptimality.thm} 
Suppose that the typical model is an FSM. Let the atypicality detector
be given an anomalous sequence generated by an FSM distinct from the
typical FSM. Then as the length of the sequence $l\to\infty$, the
probability of detecting the anomaly converges to 1, while the probability
of false alarm converges to 0. 
\end{thm}
As far as we know, nothing similar has been proven for any other anomaly
detection methods.

\subsection{\label{AnomolousGraph.sec}Anomalous Graph Detection}

In anomalous graph detection, we are given a set of training graphs
$G_{1},\ldots,G_{T}$, and the problem is then to determine if a given
graph $G$ is anomalous based on the training. We will apply atypicality
to this problem. The methodology follows directly from Definition
\ref{atypdef.thm}. We learn coding of typical graphs, Section \ref{Calc.sec},
and compare this with applying a universal source coder to $G$. In this paper, we consider unweighted, undirected graphs.

For \ER\ graphs, atypicality reduces to a hypothesis
test of $\hat{p}=p$ versus $\hat{p}\neq p$, which is of the form
$|\hat{p}-p|\geq\tau$ for some threshold. There is no reason to use
coding, and even coding structure as in \cite{ChoiSzpankowski12}
does not help: in a test of $\hat{p}=p$ versus $\hat{p}\neq p$ ,
the structural decomposition would be the same, only the coding of
the resulting bitstreams would be different.

For more complicated classes of random graphs such \BA\ or Watts Strogatz \cite{watts1998collective}, more information can be obtained using the coding algorithms developed in Section \ref{Coding.sec}. The general procedure is as follows
\begin{enumerate}
\item On the set of training graphs, we run all the coding algorithms. For
each we learn the values of the parameters (e.g., the histogram) for
the algorithm. We choose the coder that gives the shortest codelength.
The typical coder is now that algorithm with the learned parameters.
Both coder and decoder know the values of the parameters, so this
does not need to be encoded. 

\item On the set of test graphs, we run first the typical coder and obtain the typical code length $L_T$. We then
run all the coding algorithms from Section \ref{Coding.sec}; to each
codelength we have to add the overhead of encoding the parameters
(e.g., histogram). The atypical codelength, $L_A$, is now the minimum of these
codelengths, plus a few bits to tell which coder was the shortest.
The atypicality measure is  the difference between the atypical
codelength and the typical codelength, $L_A-L_T$. If $L_A-L_T < 0$, or is smaller than some threshold\footnote{The threshold has a coding interpretation: it is the number of bits required to tell the decoder an atypical coder is used \cite{HostSabetiWalton15}}, then following Definition~\ref{atypdef.thm}, the graph is declared atypical (anomalous).
\end{enumerate}
We tested this procedure by generating various random graphs with 
$n=100$ nodes.

The typical graphs were generated by using \BA \ graphs model ($m=10$). We
trained on 100 randomly generated graphs. We then generated
500 test graphs each of: 
\begin{enumerate}
\item \BA\ graphs ($m=10$) (i.e., typical graphs)
\item \BA\ graphs ($m=9$)
\item \ER graphs ($p=0.182$),  chosen so that the graph has the same average degree as the typical graph.
\item Mixture graph: combination of \BA\ graphs ($m=10$) and \ER graphs with $p=0.5$; these are essentially
\BA\ graphs with extra edges added ($p$) to make more triangles.

\end{enumerate}

%
%
%

\begin{figure}[tbh]
\includegraphics[width=3.5in]{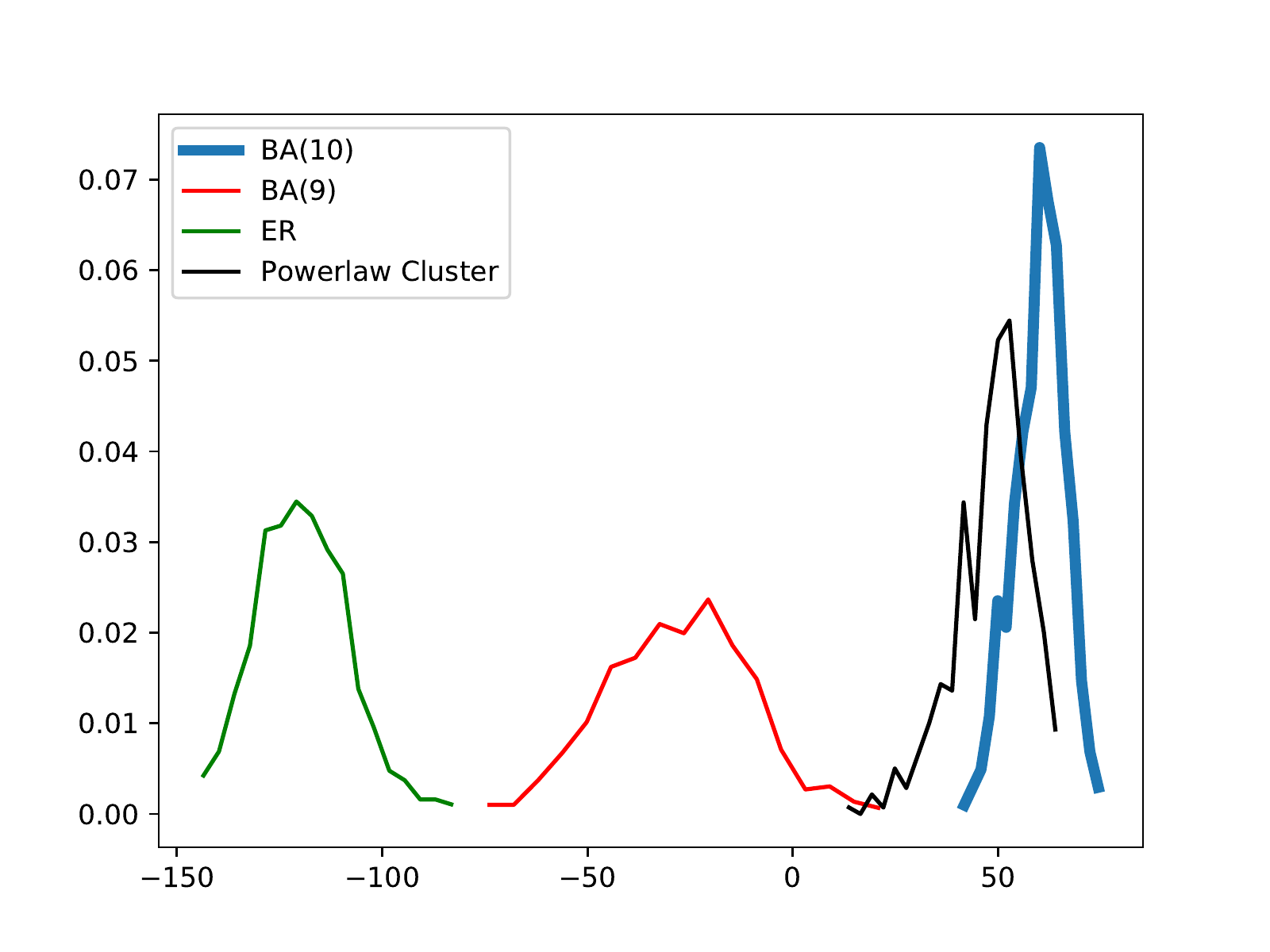}

\caption{\label{Anomaly1.fig}Pdf of atypicality measure for different types
of graphs ($n=100$). The typical graphs are BA(10), which are used
for training.}

\end{figure}

\begin{figure}[tbh]
\includegraphics[width=3.5in]{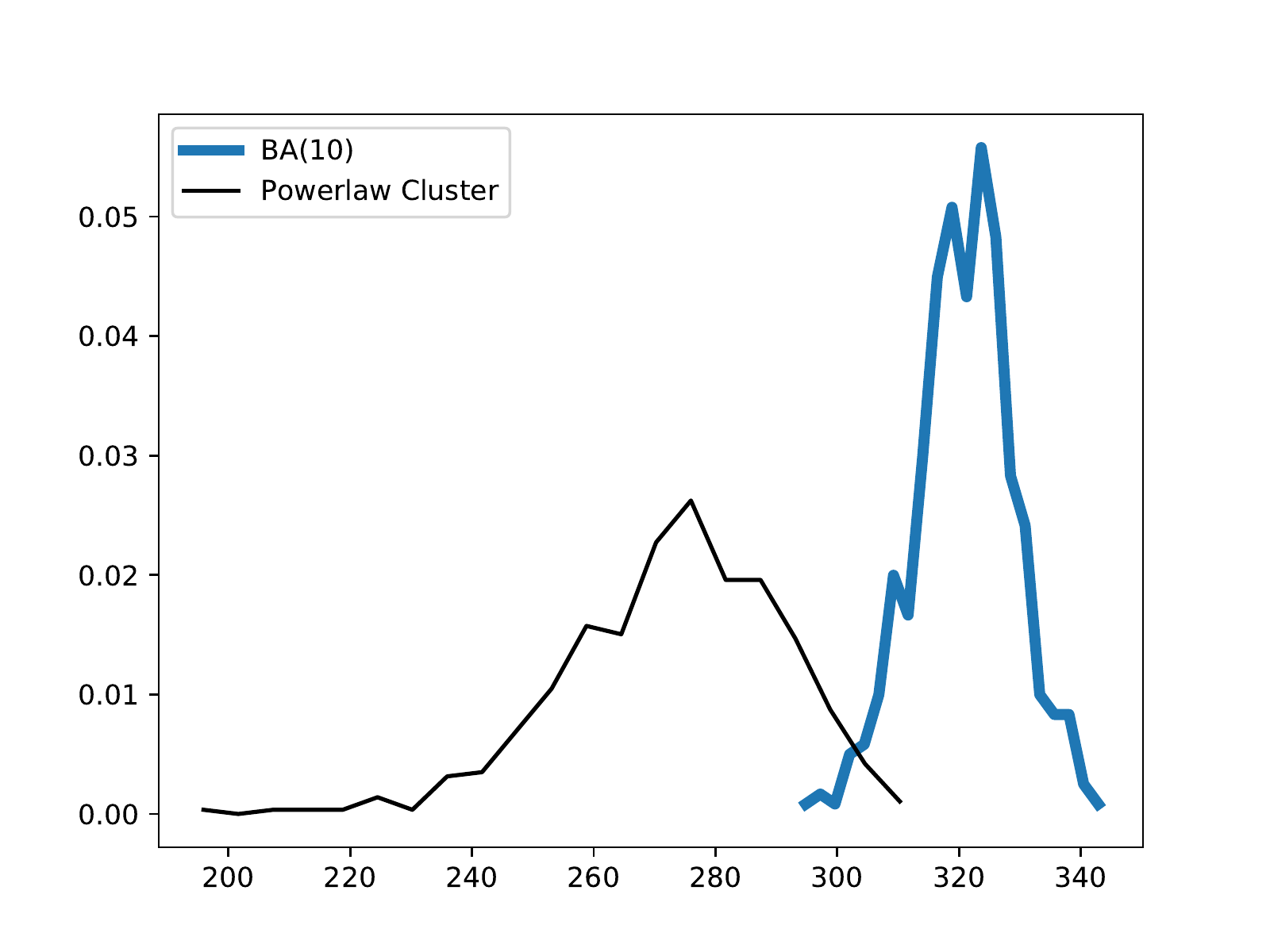}

\caption{\label{Anomaly2.fig}Pdf of atypicality measure for different types
of graphs ($n=400$). The typical graphs are BA(10), which are used
for training.}
\end{figure}

We then estimated the probability density function (pdf) of the atypicality measure:
$L_A-L_T$. The results are in Fig.
\ref{Anomaly1.fig}. We can see that \ER and \BA ($m=9$) test graphs can be easily distinguished from the typical graphs, \BA\ ($m=10$). Identifying mixture graph from \BA\ ($m=10$) is more difficult. However, due to the law of large numbers, anomaly detection improves as graph size increases. Figure~\ref{Anomaly2.fig} shows the estimated pdf of atypicality measures between mixture graph and \BA\ ($m=10$) for graphs with $n=400$ nodes; if we choose the threshold to be 305, we get $P_{\text{false alarm}}=P_{\text{miss}}=2.4\%$.


\section{Conclusions and Future Work}

In this paper we have developed a number of new universal graph coding
algorithms. The minimum codelength is found by coding with each algorithm,
and then finding the minimum (or weighting as in \cite{VolfWillems98}).
However, this still far from the state of the art for sequences, where
there a single algorithms such as CTW \cite{WillemsAl95} and Lempel-Ziv \cite{CoverBook}
that can code sequences with variable complexity. One possibility
is to generalize the triangle coding to consider structures of variable
complexity, and weight these in an approach similar to CTW.

We have shown that the coding algorithms can be used for graph anomaly
detection based on structure alone. We will consider a number of extensions. First, in most
graph-based anomaly detection problems, the anomaly is in the data
on the graph. Our idea is to combine graph structure coding with
coding of the data to get a single measure that takes into account
both data and structure. Second, we need to be able to consider graphs
of variable size; the complication here is that statistics might very
well depend on size. Finally, we will consider detecting anomalous
subgraphs.

\end{document}